\documentclass[12pt]{article}
\usepackage{amssymb}
\usepackage{amsmath}
\usepackage{bbm}
\usepackage{graphicx}

\usepackage{cite}

\numberwithin{equation}{section}

\newcommand{\be}{\begin{equation}}
\newcommand{\ee}{\end{equation}}
\newcommand{\non}{\nonumber}

\newcommand{\A}{\mathbb{A}}
\newcommand{\B}{\mathbb{B}}
\newcommand{\C}{\mathbb{C}}
\newcommand{\D}{\mathbb{D}}

\newcommand{\R}{\mathbb{R}}
\newcommand{\T}{\mathbb{T}}

\newcommand{\tb}{\mathbbm{t}}
\newcommand{\tr}{\mathop{\rm tr}\nolimits}

\begin{document}

\begin{titlepage}
\strut\hfill UMTG--309
\vspace{.5in}
\begin{center}

\LARGE Bethe ansatz on a quantum computer?\\ 
\vspace{1in}
\large Rafael I. Nepomechie\footnote{nepomechie@miami.edu}\\
Physics Department, P.O. Box 248046\\
University of Miami, Coral Gables, FL 33124\\[0.8in]
\end{center}

\vspace{.5in}

\begin{abstract}
We consider the feasibility of studying the anisotropic Heisenberg quantum spin
chain with the Variational Quantum Eigensolver (VQE) algorithm, by
treating Bethe states as variational states, and Bethe roots as
variational parameters.  For short chains, we construct exact
one-magnon trial states that are functions of the variational parameter, and
implement the VQE calculations in Qiskit.  However, exact multi-magnon
trial states appear to be out out of reach.
	
\end{abstract}

\end{titlepage}

\setcounter{footnote}{0}

\section{Introduction}\label{sec:intro}

Significant attention is being focused on an algorithm in quantum 
computing known as the Variational Quantum Eigensolver (VQE) \cite{Peruzzo_2014, McClean_2016,
alexeev2020quantum} . In the era of Noisy Intermediate-Scale
Quantum devices \cite{Preskill_2018}, this algorithm may have practical
applications in quantum chemistry \cite{Kandala_2017, 
McArdle2020QuantumCC, Arute2020}.  The
basic idea is to reduce the chemistry problem to a one-dimensional
quantum spin chain Hamiltonian, whose ground-state energy is then
estimated using the variational principle by means of a hybrid
quantum/classical computation.  A nice feature of this approach is
that the Hamiltonian does not need to be implemented in a circuit.
The main challenge is to find a suitable variational state, which
typically depends on many variational parameters.

It has long been known that certain one-dimensional quantum spin
chains are integrable, and can therefore be ``solved exactly'' by
Bethe ansatz \cite{Bethe:1931hc}.  Indeed, the eigenstates have a
general form (Bethe states) that depend on a set of parameters (Bethe
roots), which obey a set of equations called Bethe equations;
however, these equations are in general hard to solve.
Because of the relative simplicity of such Bethe-ansatz-solvable spin
chains compared with those arising in quantum chemistry, the former
could serve as useful testbeds of the VQE approach. (Simpler 
integrable models have been studied on quantum computers
using other methods, see e.g. \cite{Cervera_Lierta_2018} and references therein.)

We consider here the feasibility of studying Bethe-ansatz-solvable spin
chains with VQE, by treating Bethe states as variational states, and
Bethe roots as variational parameters.  Attractive features of this
idea are that the variational states can become exact, and that the
number of Bethe roots scales linearly with the length of the chain.

We find that such an approach is indeed feasible for short chains, 
which however can be easily solved by elementary means. But for 
longer chains, this approach does not seem feasible, due to the difficulty of 
constructing the variational states.

The outline of this paper is as follows. In Sec. \ref{sec:XXZ}, we 
introduce the model, and set up a variational problem using 
one-magnon trial states. We discuss
the cases of 2 and 4 sites, including their implementations in Qiskit 
\cite{jay_gambetta, Qiskit-Textbook} using the 
IBM Quantum Experience \cite{IBMQ}, 
in Secs. \ref{sec:N2} and \ref{sec:N4}, respectively. We briefly 
discuss multi-magnon states in Sec. \ref{sec:multimag}, and draw some 
conclusions in Sec. \ref{sec:conclusion}. The needed Bethe ansatz 
results are collected in Appendix \ref{sec:BA}.

\section{The model}\label{sec:XXZ}

We consider as our working example the periodic spin-1/2 anisotropic
Heisenberg (XXZ) quantum spin chain with $N$ sites in the ferromagnetic regime, 
whose Hamiltonian is given by
\be
{\cal H} = -\frac{1}{4}\sum_{k=1}^{N} \left[
\sigma^{x}_{k} \sigma^{x}_{k+1} + \sigma^{y}_{k} \sigma^{y}_{k+1} +
\cosh(\eta) \left( \sigma^{z}_{k} \sigma^{z}_{k+1} - \sigma^{0}_{k} 
\sigma^{0}_{k+1} \right)\right]\,,
\qquad \sigma^{i}_{N+1} \equiv \sigma^{i}_{1}\,,
\label{Hamiltonian}
\ee
where $\sigma^{i}_{k}$ denotes the $i^{th}$ Pauli matrix at site $k$ 
($\sigma^{0}$ is the two-dimensional identity matrix), and $\eta>0$ is the 
anisotropy parameter. This 
model and its cousins have a long history, and have applications ranging from condensed 
matter physics and statistical mechanics to high-energy theory \cite{Batchelor:2007}.

This model has a $U(1)$ symmetry
\be
\left[ {\cal H} \,, S^{z} \right] = 0 \,, \qquad 
S^{z} = \frac{1}{2} \sum_{k=1}^{N} \sigma^{z}_{k}\,,
\label{Sz}
\ee
as well as charge conjugation symmetry
\be
{\cal C}\, {\cal H}\, {\cal C} = {\cal H} \,, \qquad
{\cal C} = \prod_{k=1}^{N} \sigma^{x}_{k} \,.
\label{ccsymetry}
\ee
Note that ${\cal C}\, S^{z}\, {\cal C} = -S^{z}$.
Hence, if $|\Psi\rangle$ is a simultaneous eigenstate of ${\cal H}$ 
and $S^{z}$ with nonzero eigenvalue of $S^{z}$, then it is 
degenerate, since ${\cal C}\, |\Psi\rangle$ has the same energy as (and 
is linearly independent from) $|\Psi\rangle$.

Let $|\Psi_{0}\rangle$ denote the reference state
\be
|\Psi_{0}\rangle = {1\choose 0}^{\otimes N} \,.
\label{reference}
\ee
Both $|\Psi_{0}\rangle$ and ${\cal C} |\Psi_{0}\rangle$ are ground 
states of the Hamiltonian (\ref{Hamiltonian}), with energy 0 and 
$S^{z}$ eigenvalues $N/2$ and $-N/2$ respectively. 

Our initial (admittedly, modest) objective is to use VQE to estimate the 
energy of the first excited state. The normalized trial state 
$|\Psi\rangle$  must be orthogonal to both ground states
\be
\langle \Psi_{0} | \Psi\rangle = \langle \Psi_{0} | {\cal C}  
| \Psi\rangle = 0 \,.
\label{requirement}
\ee
Indeed, the usual variational argument can then be applied: let 
$\{ |n\rangle \}$ be a complete orthonormal set of eigenstates of the 
Hamiltonian (\ref{Hamiltonian})
\be
{\cal H}\, |n\rangle = E_{n}\,  |n\rangle\,, \qquad \langle n | m 
\rangle = \delta_{n,m} \,, \qquad \sum_{n=0}^{2^{N}-1} | n \rangle  
\langle n |  = 1 
\,,
\ee
with ordered energies $E_{n} \le E_{n+1}$. The requirement 
(\ref{requirement}) translates to 
\be
\langle n | \Psi \rangle = 0 \,, \qquad n = 0, 1 \,.
\label{requirement1}
\ee
We then have
\be
1 = \langle \Psi | \Psi \rangle = \sum_{n=0}^{2^{N}-1} 
| \langle n |  \Psi \rangle |^{2} 
= \sum_{n=2}^{2^{N}-1} 
| \langle n |  \Psi \rangle |^{2} 
\ee
and
\begin{align}
\langle \Psi | {\cal H} | \Psi \rangle &= 
\sum_{m, n=0}^{2^{N}-1} 
\langle \Psi | m \rangle \langle m| {\cal H} | n \rangle
\langle n | \Psi \rangle 
= \sum_{n=0}^{2^{N}-1} E_{n} | \langle n |  \Psi \rangle |^{2}  
= \sum_{n=2}^{2^{N}-1} E_{n} | \langle n |  \Psi \rangle |^{2} \non \\
&= E_{2} + \sum_{n=3}^{2^{N}-1} (E_{n} - E_{2}) | \langle n |  \Psi \rangle |^{2} 
\ge E_{2} \,.
\end{align}
That is, the expectation value of the Hamiltonian in the trial state 
is greater than or equal to the energy of the first excited state.

We take as our trial state the so-called one-magnon Bethe state
\be
| \Psi \rangle = \frac{| \psi \rangle}{\sqrt{\langle \psi | \psi 
\rangle} }\,,  \qquad
| \psi \rangle = B(p) |\Psi_{0}\rangle\,,  
\label{trialstate}
\ee
where the operator $B(p)$, which is \emph{not} unitary, is defined in Appendix \ref{sec:BA}. The 
requirement (\ref{requirement}) can be shown to be satisfied (for $N>1$) by using 
the facts
\be
\langle \Psi_{0} | B(p)  = 0 
\label{fact1}
\ee
and 
\be
\left[ S^{z} \,, B(p) \right] = - B(p) \,.
\label{SzB}
\ee
Indeed, from (\ref{fact1}), we see that
\be
\langle \Psi_{0} | \psi \rangle  = \langle \Psi_{0} | B(p) |\Psi_{0}  
\rangle = 0 \,;
\label{res1}
\ee
and from (\ref{SzB}), we obtain
\be
\langle \Psi_{0} | {\cal C} \left[ S^{z} \,, B(p) \right] |\Psi_{0}  \rangle
= - \langle \Psi_{0} | {\cal C} B(p) |\Psi_{0} \rangle \,,
\label{bis0}
\ee
whose left-hand-side can be evaluated as follows:
\begin{align}
\langle \Psi_{0} | {\cal C} \left[ S^{z} \,, B(p) \right] |\Psi_{0}  
\rangle & = \langle \Psi_{0} | \left( {\cal C} S^{z} B(p) -  {\cal C} 
B(p) S^{z} \right) |\Psi_{0} \rangle \\
& = \langle \Psi_{0} | \left( - S^{z} {\cal C}  B(p) -  {\cal C} 
B(p) S^{z} \right) |\Psi_{0} \rangle \label{bis1}\\
& = - N \langle \Psi_{0} | {\cal C} B(p) |\Psi_{0} \rangle 
\label{bis2} \,.
\end{align}
In passing to (\ref{bis1}), we have used the fact (noted below (\ref{ccsymetry})) that ${\cal C}\, S^{z}\, {\cal 
C} = -S^{z}$; and in passing to (\ref{bis2}), we have 
used the fact $S^{z}  |\Psi_{0} \rangle = (N/2) |\Psi_{0} \rangle$,
which follows from the remark below (\ref{BAenergy}). In view of (\ref{bis0}), we 
see that 
\be
(N-1) \langle \Psi_{0} | {\cal C} B(p) |\Psi_{0} \rangle = 0 \,.
\ee
Hence, for $N>1$, we conclude that
\be
\langle \Psi_{0} | {\cal C} | \psi \rangle  = \langle \Psi_{0} | {\cal C} B(p) |\Psi_{0} \rangle = 0 \,.
\label{res2}
\ee
Eqs. (\ref{res1}) and (\ref{res2}) imply that the trial state 
(\ref{trialstate}) is orthogonal to both ground 
states. (For other approaches to studying excited 
states using VQE, see e.g. \cite{McClean:2017, Higgott:2019}.)

Note that, as a consequence of (\ref{SzB}), the trial state 
(\ref{trialstate}) has $S^{z}$ eigenvalue $(N/2)-1$.
Note also that the trial state depends on a single variational
parameter $p$, which we take to be real. The cases $N=2$ 
and $N=4$ are explicitly worked out in Secs. \ref{sec:N2} and \ref{sec:N4}, 
respectively.
 
\section{$N=2$}\label{sec:N2}
 
We begin with the simplest case of 2 sites, for which there are 4
states: the two zero-energy ground states $|\Psi_{0}\rangle$ and
${\cal C} |\Psi_{0}\rangle$, and the first and second excited states.
By direct diagonalization of the Hamiltonian (\ref{Hamiltonian}), one
finds that the excited states have energies $\cosh(\eta) - 1$ and
$\cosh(\eta) + 1$, respectively.  We shall see that in fact
\emph{both} excited-state energies can be obtained using VQE.

The trial state (\ref{trialstate}) is given for $N=2$ by
\be
| \Psi \rangle = \frac{1}{\sqrt{2}}(0, e^{i p}, 1, 0) \,,
\label{trialstate2}
\ee
up to an overall phase.
In order to implement VQE, we must first re-express this state as a
product of standard 1-qubit and 2-qubit unitary gates acting
on the reference state $|\Psi_{0}\rangle$. Up to an overall phase,
we find that\footnote{Following the common convention in the quantum 
computing literature, we label the 2-dimensional vector spaces $V$ from right to left, 
starting from 0: $\stackrel{\stackrel{1}{\downarrow}}{V} \otimes
\stackrel{\stackrel{0}{\downarrow}}{V}$; and the 1-qubit operator $O$ 
acting on vector space $i$ is denoted by $O_{i}$. The permutation (SWAP) matrix
\be
{\cal P} = \begin{pmatrix}
1 &0 &0 &0 \\
0 &0 &1 &0 \\
0 &1 &0 &0 \\
0 &0 &0 &1 
\end{pmatrix}  
\label{permutation}
\ee
can be used to change vectors spaces: $O_{j} = {\cal P}_{ij}\, 
O_{i}\, {\cal P}_{ij}$.
}
\be
|\Psi \rangle = X_{0}\, C_{10}\, U^{(3)}_{1}(\tfrac{\pi}{2},-p, 0)\, 
|\Psi_{0}\rangle \,,
\label{N2result}
\ee
where $X = \sigma^{x}$, $C_{ij}$ is a CNOT gate with control qubit $i$ and 
target qubit $j$
\be
C_{10} = \begin{pmatrix}
1 &0 &0 &0 \\
0 &1 &0 &0 \\
0 &0 &0 &1 \\
0 &0 &1 &0 
\end{pmatrix} \,,
\label{C10}
\ee
and $U^{(3)}(\theta,\phi, \lambda)$ is defined by
\be
U^{(3)}(\theta,\phi, \lambda) =
\begin{pmatrix}
\cos(\theta/2) &-e^{i \lambda} \sin(\theta/2) \\
e^{i \phi} \sin(\theta/2) & e^{i \phi} e^{i \lambda}\cos(\theta/2)
\end{pmatrix} \,.
\label{U3}
\ee

The exact energy of the first-excited state can be obtained by minimizing 
$\langle \Psi | {\cal H} | \Psi \rangle = \cosh(\eta) - \cos(p)$; the corresponding optimal value of 
the variational parameter is $p= 0 \,\, (\text{mod}\,\, 2\pi)$, 
independently of the value of $\eta$.  

For $N=2$, the second-excited state also has the
one-magnon form (\ref{trialstate2}), see Appendix \ref{sec:BA}. Since this state is in fact the state
with \emph{highest} energy, we can obtain its 
energy via VQE by minimizing $\langle \Psi | (-{\cal H}) | \Psi
\rangle  = \cos(p) -\cosh(\eta) $, and then taking the absolute value of the result. The 
corresponding optimal value of 
the variational parameter in this case is $p= \pi \,\, (\text{mod}\,\, 2\pi)$.

\subsection{Qiskit implementation}

\begin{figure}[htb]
\centering
	\includegraphics[width=8cm]{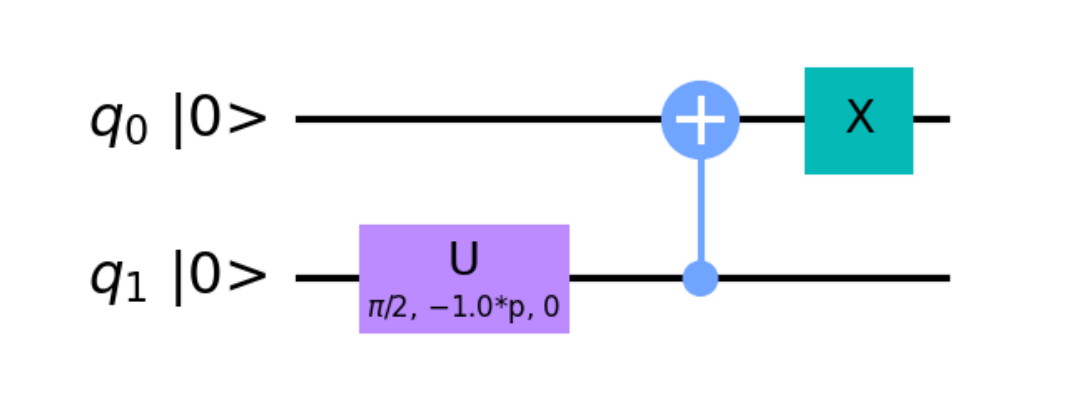}
\caption{Circuit for the trial state $|\Psi \rangle$ given by 
(\ref{N2result})}
\label{fig:N2}
\end{figure}

We implemented these variational computations in Qiskit using its VQE
algorithm, which iteratively evaluated $\langle \Psi | {\cal H} | 
\Psi \rangle$, using a classical optimizer to update the value of $p$.
We inputted the Hamiltonian (\ref{Hamiltonian}) as a 2-qubit 
SummedOp, and used the Constrained Optimization By Linear Approximation
(COBYLA) as the classical optimizer (maxiter=10).
The circuit diagram for 
the trial state $|\Psi \rangle$  given by (\ref{N2result}) is shown in Fig. \ref{fig:N2}.
For concreteness, we set $\eta=1$.
Typical results, together with the exact result, are presented in 
Table \ref{table:N2}. The results from the real device 
\verb|ibmq_valencia| (5 qubits, QV16), using measurement error mitigation,
have error of approximately $1\%$.

\begin{table}[h!]
\centering
\begin{tabular}{|l|c|c||c|c|}
\hline
Backend & $E_{\text{first-excited state}}$ & $p \,\, (\text{mod}\,\, 2\pi)$ & 
$E_{\text{second-excited state}}$ & $p \,\, (\text{mod}\,\, 2\pi)$\\   
\hline
Exact &  0.54308063 & 0 & 2.54308063  & 3.14159265 \\
Statevector simulator & 0.54316139 & -0.01270853 & 2.54203089 & -3.09576858 \\
Qasm simulator ($10^{3}$ shots, no noise) & 0.54308063 & -0.01549807 & 2.54308063 & 
3.17170765 \\
Real 5-qubit device ($10^{3}$ shots) & 0.54549027 & -0.17336981 & 2.54191949 & 2.97042731 \\
\hline
\end{tabular}
\caption{VQE energies and optimal parameter values for the 
Hamiltonian (\ref{Hamiltonian}) with 
$N=2$ and $\eta=1$, using the trial state depicted in Fig. \ref{fig:N2}}
\label{table:N2}
\end{table}

\section{$N=4$}\label{sec:N4}

We now consider the case of 4 sites. By direct diagonalization 
of the Hamiltonian (\ref{Hamiltonian}), one finds that 
the exact energy of the (two-fold degenerate) first-excited states is $\cosh(\eta) -1$.

The trial state (\ref{trialstate}) is now given by
\be
| \Psi \rangle = (0, \tfrac{1}{2}e^{3 i p/2}, \tfrac{1}{2}e^{i p/2}, 
0, \tfrac{1}{2}e^{-i p/2}, 0, 0, 0, \tfrac{1}{2}e^{-3i p/2}, 0, 0, 0, 0, 0, 0, 0) \,,
\label{trialstate4}
\ee
up to an overall phase. The main difficulty is to re-express this state as a
product of standard 1-qubit and 2-qubit unitary gates acting
on the reference state $|\Psi_{0}\rangle$. To this end, we follow 
\cite{Plesch_2011}, 
and perform the Schmidt decomposition of (\ref{trialstate4}), thereby 
obtaining
\be
| \Psi \rangle = \left({\cal U} \otimes {\cal V}\right) C_{31}\, 
C_{20}\, H_{2}\, |\Psi_{0}\rangle \,,
\label{N4resulta}
\ee
where $H$ is the Hadamard gate
\be
H = 
\frac{1}{\sqrt{2}} \begin{pmatrix}
                            1 & 1 \\
                            1 & -1
			         \end{pmatrix}  \,, 
\ee
and ${\cal U}$ and ${\cal V}$ are the following unitary matrices
\be
{\cal U} = \begin{pmatrix}
e^{i p/2} & 0 & 0 & 0 \\
0 & \frac{1}{\sqrt{2}}e^{-i p/2} & 0 & -\frac{1}{\sqrt{2}}e^{i p} \\
0 & \frac{1}{\sqrt{2}}e^{-3i p/2} & 0 & \frac{1}{\sqrt{2}} \\
0 & 0 & 1 & 0 
\end{pmatrix}  \,, \qquad
{\cal V} = \begin{pmatrix}
0 & 1 & 0 & 0 \\
\frac{1}{\sqrt{2}}e^{i p} & 0 & 0 & -\frac{1}{\sqrt{2}}e^{i p} \\
\frac{1}{\sqrt{2}} & 0 & 0  & \frac{1}{\sqrt{2}} \\
0 & 0 & 1 & 0 
\end{pmatrix} \,.
\ee
We now need to decompose ${\cal U}$ and ${\cal V}$ into gates. To this 
end, we follow \cite{Shende_2004, Shende_2004b, j2020quantum}, and 
eventually obtain (up to irrelevant overall phases)
\begin{align}
{\cal U} &= \left[ U^{(3)}(\tfrac{\pi}{2}, -\tfrac{3p}{2}, 0) \otimes 
U^{(3)}(\tfrac{\pi}{2}, \tfrac{1}{2}(\pi-p), \pi)  \right]\, V(0, 
\tfrac{\pi}{2}, \tfrac{\pi}{4})\, \non \\
& \qquad \times  
\left[ U^{(3)}(\tfrac{\pi}{2}, -\tfrac{\pi}{4}, \tfrac{1}{2}(3p+\pi)) \otimes 
U^{(3)}(0, 0, \tfrac{1}{2}(3\pi-p))  \right] \label{N4resultb}\\
{\cal V} &= \left[ U^{(3)}(\tfrac{\pi}{2}, \tfrac{1}{2}(\pi-p), 0) \otimes 
U^{(3)}(\tfrac{\pi}{2}, \tfrac{p}{2}, -\tfrac{\pi}{2})  \right]\, 
V(\tfrac{1}{2}(3\pi-p), \tfrac{3\pi}{4}, \tfrac{3\pi}{4})\, \non \\
& \qquad \times 
\left[ U^{(3)}(\tfrac{\pi}{2}, -\tfrac{\pi}{2}, \tfrac{\pi}{2}) \otimes 
U^{(3)}(\tfrac{\pi}{2}, -\tfrac{\pi}{2}, -\pi)  \right] \,,
\label{N4resultc}
\end{align}
where $V(\alpha, \beta, \delta)$ is the 2-qubit operator defined as
\be
V(\alpha, \beta, \delta) = Z_{1}\, C_{01}\, U^{(3)}_{0}(-\alpha,\pi,\pi)\, 
C_{10}\, U^{(3)}_{1}(0,0,\delta)\, U^{(3)}_{0}(-\beta,\pi,\pi)\,
C_{01}\, U^{(3)}_{0}(0,0, -\frac{\pi}{2}) \,,
\label{N4resultd}
\ee
with $Z = \sigma^{z}$.

The exact energy of the first-excited states can be obtained by
minimizing $\langle \Psi | {\cal H} | \Psi \rangle = \cosh(\eta) -
\cos^{3}(p)$; the corresponding optimal value of the variational
parameter is $p= 0 \,\, (\text{mod}\,\, 2\pi)$.

\subsection{Qiskit implementation}

\begin{figure}[htb]
\centering
	\includegraphics[width=8cm]{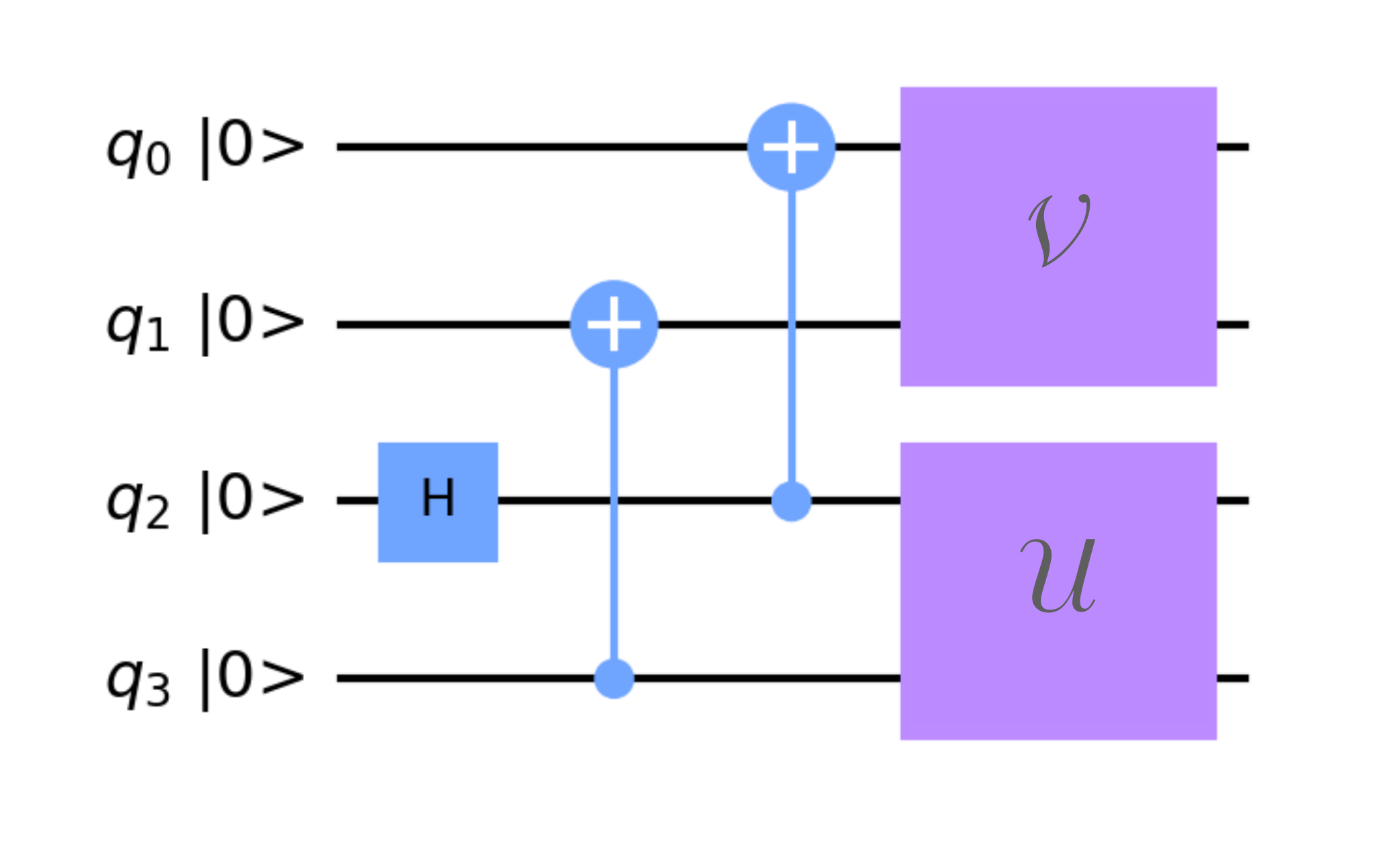}
\caption{Circuit for the trial state $|\Psi \rangle$ given by 
(\ref{N4resulta})}
\label{fig:psi}
\end{figure}

\begin{figure}[htb]
\centering
	\includegraphics[width=16cm]{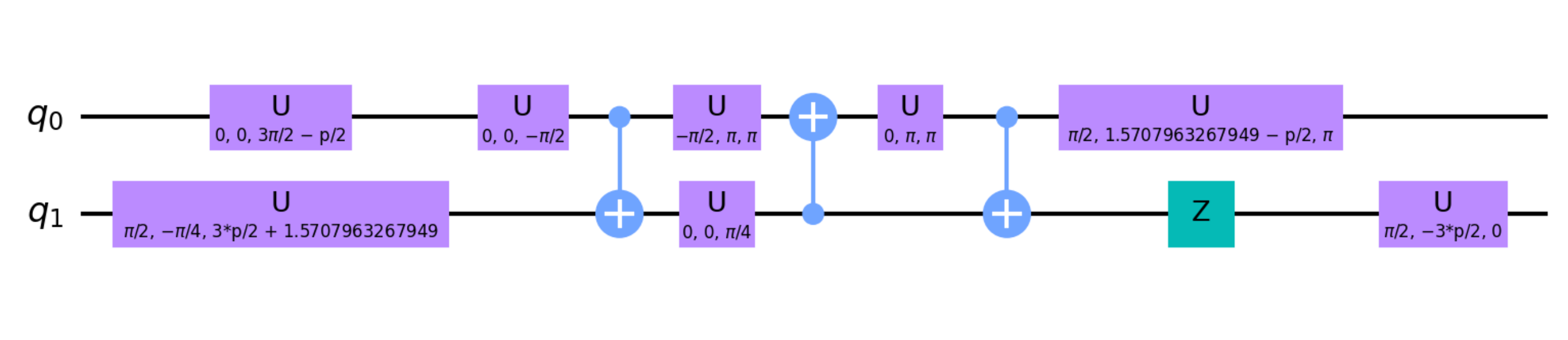}
\caption{Circuit for the matrix ${\cal U}$ given by (\ref{N4resultb})}
\label{fig:calU}
\end{figure}

\begin{figure}[htb]
\centering
	\includegraphics[width=16cm]{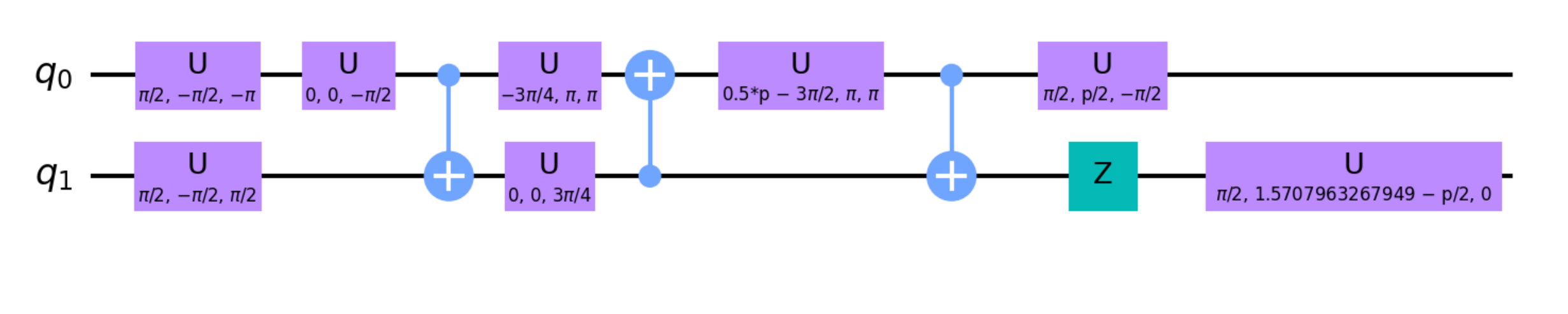}
\caption{Circuit for the matrix ${\cal V}$ given by (\ref{N4resultc})}
\label{fig:calV}
\end{figure}

We have again implemented the variational computation in Qiskit using its VQE
algorithm with the COBYLA classical optimizer (maxiter=20), inputting the Hamiltonian 
(\ref{Hamiltonian}) as a 4-qubit \mbox{SummedOp}. We again set $\eta=1$.
Circuit diagrams for the trial state $|\Psi \rangle$ 
(\ref{N4resulta}), as well as for the matrices
${\cal U}$ (\ref{N4resultb}) and ${\cal V}$ (\ref{N4resultc}), are shown in Figs.
\ref{fig:psi}, \ref{fig:calU} and \ref{fig:calV},  
respectively.
Typical results, together with the exact result, are presented in 
Table \ref{table:N4}. The results from the real device 
\verb|ibmq_santiago| (5 qubits, QV32), using measurement error mitigation,
have error of approximately $35\%$.

\begin{table}[h!]
\centering
\begin{tabular}{|l|c|c|}
\hline
Backend & $E_{\text{first-excited states}}$ & $p \,\, (\text{mod}\,\, 2\pi)$\\   
\hline
Exact &  0.54308063 & 0 \\
Statevector simulator & 0.54308065 & -0.00008658 \\
Qasm simulator ($8192$ shots, no noise) & 0.53905231 & 0.05466209 \\
Real 5-qubit device ($8192$ shots) & 0.73343292  & 0.13868187\\
\hline
\end{tabular}
\caption{VQE energy and optimal parameter value for the 
Hamiltonian (\ref{Hamiltonian}) with $N=4$ and $\eta=1$, 
using the trial state depicted in Figs. \ref{fig:psi}-\ref{fig:calV}}
\label{table:N4}
\end{table}

\section{Multi-magnon states}\label{sec:multimag}

We have thus far focused on a special class of eigenstates of the 
Hamiltonian (\ref{Hamiltonian}) described by
1-magnon states (\ref{trialstate}), with $p$ real. However, in order to describe all 
the eigenstates, it is necessary to consider $M$-magnon states, with 
$M=0, 1, \ldots, \lfloor N/2 \rfloor$, and with complex-valued $p$'s.
In particular, already for $N=4$, we need also 
2-magnon states. Restricting to real values of $p_{1}$ and $p_{2}$ 
(which is the case for the ground state of $-{\cal H}$), one can show that
\be
B(p_{1})\, B(p_{2}) |\Psi_{0}\rangle  \propto (0,0,0, e^{i(p_{1}+p_{2})}, 0, 
\xi, 1, 0, 0, \zeta, \xi^{*}, 0, e^{-i(p_{1}+p_{2})}, 0, 0, 0)  \,,
\ee 
where
\begin{align}
\xi &= \frac{1}{1+e^{-i(p_{1}+p_{2})}}\left[e^{i p_{1}} + e^{-i p_{1}} +  e^{i p_{2}} + e^{-i p_{2}}  
- 2\cosh(\eta) \right] \,, \non \\
\zeta &= \frac{1}{1+e^{i(p_{1}+p_{2})}}\left[
1 + e^{2 i p_{1}} + e^{2 i p_{2}} + e^{i (p_{1} - p_{2})} + e^{i 
(p_{2} - p_{1})} + e^{i (p_{1} + p_{2})} - 2 \cosh(\eta)(e^{i p_{1}} 
+ e^{i p_{2}} ) \right] \,,
\end{align}
and $\xi^{*}$ denotes the complex conjugate of $\xi$.
Unfortunately, we are unable 
to find an explicit gate decomposition of the corresponding normalized state, 
since the Schmidt 
decomposition involves square roots of complicated functions of 
$p_{1}$, $p_{2}$ and $\eta$.  Evidently, exact multi-magnon trial 
states as explicit functions of the variational parameters
are out of reach.

\section{Conclusions}\label{sec:conclusion}

We have explicitly derived one-magnon trial states describing the 
first-excited states of the Hamiltonian (\ref{Hamiltonian}) for the cases
$N=2$ and $N=4$, see (\ref{N2result}) and
(\ref{N4resulta})-(\ref{N4resultd}), respectively.  The main lesson
from these computations is that it is possible to obtain 
gate decompositions of exact one-magnon states that are explicit functions of a 
variational parameter $p$.  We expect that, with additional effort, similar gate
decompositions of exact one-magnon states could also
be derived for higher values of $N$.  However, the corresponding 
energies of this special class of
states can be obtained much more easily by alternative means.

The full power of Bethe ansatz comes from the fact that all 
eigenstates can be described by multi-magnon states. 
In principle, this fact could be exploited within the VQE context by using multi-magnon trial 
states to find the minimum eigenvalue of the antiferromagnetic Hamiltonian $-{\cal H}$, whose ground state lies in the sector 
with $N/2$ magnons.\footnote{Actually, the ground-state energy of $-{\cal H}$ can 
be accurately estimated even for relatively large values of $N$ (say, 
$\sim 10^{3}$), 
since there are effective methods for numerically solving the
Bethe equations (\ref{BAE}) for the special case when all the roots 
$v_{1}, \ldots, v_{N/2}$ are real.} 
For $N=2$, the antiferromagnetic ground state is a 1-magnon state, which was already studied in Sec. \ref{sec:N2}.
For $N\ge 4$, the antiferromagnetic ground state lies in a sector with 2 or more magnons.
However, as argued in Sec. \ref{sec:multimag}, exact multi-magnon trial states appear to be 
out of reach. We conclude that a VQE approach based on exact multi-magnon trial states
that are explicit functions of the variational parameters does not seem feasible. 

One alternative strategy could be to determine the gate decomposition
of the exact multi-magnon trial state dynamically. That is, to 
(classically) recompute the gate decomposition
at each stage of the VQE iteration, where the variational parameters
have explicit numerical values. Another alternative strategy could be to 
look for good approximate multi-magnon trial states. We leave these and 
related questions for future investigations.

\section*{Acknowledgments}

I acknowledge use of the IBM Quantum Experience for this work.
I thank Stephan Eidenbenz for helpful correspondence.

\appendix

\section{Bethe ansatz}\label{sec:BA}

We collect here the basic results from algebraic Bethe ansatz that 
are needed to construct the Hamiltonian and trial states.  For further details,
see e.g. \cite{Korepin:1993, Faddeev:1996iy, Nepomechie:1998jf}

We start from the (shifted) R-matrix $\R(v)$ acting on $V\otimes V$
\be
\R(v) = \begin{pmatrix}
\sin(v+\tfrac{i \eta}{2}) & 0 & 0 & 0 \\
0 & \sin(v-\tfrac{i \eta}{2}) & i \sinh(\eta) & 0  \\
0 & i \sinh(\eta) & \sin(v-\tfrac{i \eta}{2}) & 0  \\
0 & 0 & 0 & \sin(v+\tfrac{i \eta}{2})   
\end{pmatrix} \,,
\ee
which is proportional to the permutation matrix 
(\ref{permutation}) when $v=\tfrac{i \eta}{2}$, and
which is a solution of the (shifted) Yang-Baxter equation on $V\otimes V\otimes 
V$ \footnote{We follow in the appendix (in contrast with the body of 
the paper) the standard convention in the Bethe ansatz 
literature, and label the 2-dimensional vector spaces $V$ from left 
to right, starting from 1: $\stackrel{\stackrel{1}{\downarrow}}{V} \otimes
\stackrel{\stackrel{2}{\downarrow}}{V}$.}
\be
\R_{12}(v_{1} - v_{2}+\tfrac{i \eta}{2})\,  \R_{13}(v_{1})\, \R_{23}(v_{2}) 
= \R_{23}(v_{2})\,  
\R_{13}(v_{1})\, \R_{12}(v_{1} - v_{2}+\tfrac{i \eta}{2}) \,.
\label{YBE}
\ee
We define the monodromy matrix $\T_{0}(v)$ by
\be
\T_{0}(v) = \R_{0 N}(v)\, \ldots \R_{0 1}(v)\,.
\label{monodromy}
\ee
This is a matrix acting on $V^{\otimes (N+1)}$;
the vector space 0 is called ``auxiliary'', and the vector spaces $1, 
\ldots, N$ are called ``quantum''; it is customary to suppress the 
quantum subscripts of the monodromy matrix. By tracing over the 
auxiliary space, we arrive at the transfer matrix $\tb(v)$
\be
\tb(v) = \tr_{0} \T_{0}(v) \,,
\ee
which obeys (as a consequence of the Yang-Baxter equation (\ref{YBE}))
the commutativity property
\be
\left[\tb(v_{1}) \,, \tb(v_{2}) \right] = 0 \,,
\label{commutativity}
\ee
which is the hallmark of quantum integrability. The transfer matrix is the 
generating function of the Hamiltonian (\ref{Hamiltonian}) 
\be
{\cal H} = 
-\frac{i}{2}\sinh(\eta)\frac{d}{dv}\log\left(\tb(v)\right)\Big\vert_{v=\frac{i 
\eta}{2}}+ \frac{1}{2} N \cosh(\eta) 
\ee
and of higher conserved commuting quantities.

To construct the eigenstates, we write the monodromy matrix 
(\ref{monodromy}) as a $2 \times 2$ matrix in the auxiliary space
\be
\T_{0}(v) = \begin{pmatrix}
\A(v) & \B(v) \\
\C(v) & \D(v) 
\end{pmatrix}_{0} \,,
\ee
and use the quantum-space operator $\B(v)$ as a creation operator acting on the reference state 
(\ref{reference}). Indeed, it can be shown that the Bethe states
\be
\B(v_{1}) \ldots \B(v_{M}) |\Psi_{0}\rangle
\ee
are eigenstates of the transfer matrix (and, therefore, of the
Hamiltonian) if $v_{1}, \ldots, v_{M}$ are pairwise distinct and
satisfy the Bethe equations
\be
\left(\frac{\sin(v_{j}  + \tfrac{i \eta}{2})}
{\sin(v_{j} - \tfrac{i \eta}{2})}\right)^{N} = \prod_{k=1; k \ne j}^{M} 
\frac{\sin(v_{j}  -v_{k} + i \eta)}{\sin(v_{j}  -v_{k} - i \eta)} \,, 
\qquad j = 1, \ldots, M\,. 
\label{BAE}
\ee
We can take advantage of the charge conjugation symmetry 
(\ref{ccsymetry}) to restrict the values of $M$ to 
$M = 0, 1, \ldots, \lfloor N/2 \rfloor$. The corresponding 
eigenvalues of the Hamiltonian are given by
\be
E = \frac{1}{2}\sum_{j=1}^{M} \frac{1}{\sin(v_{j}  + \tfrac{i 
\eta}{2})\, \sin(v_{j}  - \tfrac{i \eta}{2})} \,.
\label{BAenergy}
\ee
The Bethe states are also eigenstates of $S^{z}$ (\ref{Sz}), with 
eigenvalue $(N/2)-M$. We emphasize that the Bethe equations (\ref{BAE}) 
cannot be solved analytically unless either the number of sites $N$ or the number
of magnons $M$ is very small.  In general, one can only solve these
equations numerically, and even this is generally very difficult.

It is convenient to change from the variable $v$ to the new variable 
$p$ defined by
\be
\frac{\sin(v  + \tfrac{i \eta}{2})}
{\sin(v - \tfrac{i \eta}{2})} = e^{-ip} \,,
\ee
as suggested by the LHS of the Bethe equations (\ref{BAE}); and 
we set $B(p) \equiv \B(v)$.


\providecommand{\href}[2]{#2}\begingroup\raggedright\endgroup

\end{document}